\documentclass[prl,amsmath,amsfonts,twocolumn]{revtex4-1}
\pdfoutput=1 % if your are submitting a pdflatex (i.e. if you have
\usepackage[utf8]{inputenc}

\usepackage[T1]{fontenc} % if needed

\usepackage{xcolor}
\usepackage{cancel}
\usepackage{tensor}
\usepackage{hyperref}
\usepackage{caption}
\usepackage{subcaption}
\usepackage{siunitx}
\usepackage{amsfonts}
\usepackage{amssymb}
\usepackage{slashed} %new 8/3/16
\usepackage{tcolorbox} %new 17/3/16
\usepackage{mdframed} %new 17/3/16

\begin{document}

\newcommand{\Mp}{M_{\mathrm{P}}}
\newcommand{\Mnn}{\mu_n}
\newcommand{\Mns}{\mu_\times}
\newcommand{\Mss}{\mu_s}
\newcommand{\Vv}{V_{v}}
\newcommand{\Vvv}{V_{vv}}
\newcommand{\Vvw}{V_{vw}}
\newcommand{\Vww}{V_{ww}}

\newcommand{\Nf}{{N_{\mathrm{f}}}}
\newcommand{\Lh}{\Lambda_{\mathrm{h}}}
\newcommand{\Lv}{\Lambda_{\mathrm{v}}}
\newcommand{\epsilonV}{\epsilon_{\mathrm{V}}}
\newcommand{\etaV}{\eta_{\mathrm{V}}}
\newcommand{\epsilonH}{\epsilon_{\mathrm{H}}}
\newcommand{\etaH}{\eta_{\mathrm{H}}}
\newcommand{\epsiloni}{\epsilon_{\mathrm{i}}}
\newcommand{\etai}{\eta_{\mathrm{i}}}

\newcommand{\db}[1]{\dot{\bar{#1}}}
\newcommand{\ddb}[1]{\ddot{\bar{#1}}}
\newcommand{\tsum}{{\textstyle{\sum}}}

\newcommand{\vv}[1]{\mathbf{#1}}
\newcommand{\vh}[1]{\mathbf{\hat{#1}}}
\newcommand{\nb}{\mathbf{\nabla}}
\newcommand{\dv}{\mathbf{\nabla}\cdot}
\newcommand{\cl}{\mathbf{\nabla}\times}
\newcommand{\bs}[1]{\boldsymbol{#1}}
\newcommand{\vvv}{\mathbf{v}}
\newcommand{\vvu}{\mathbf{u}}
\newcommand{\vvx}{\mathbf{x}}
\newcommand{\vvk}{\mathbf{k}} %new
\newcommand{\vvp}{\mathbf{p}} %new
\newcommand{\vvr}{\mathbf{r}}
\newcommand{\vvR}{\mathbf{R}} %new
\newcommand{\vvG}{\mathbf{G}} %new
\newcommand{\mc}[1]{\mathcal{#1}} %new
\newcommand{\vvrd}{\dot{\mathbf{r}}}
\newcommand{\vvrdd}{\ddot{\mathbf{r}}}
\newcommand{\rrd}{\dot{r}}
\newcommand{\rrdd}{\ddot{r}}
\newcommand{\thetad}{\dot{\theta}}
\newcommand{\thetadd}{\ddot{\theta}}
\newcommand{\phid}{\dot{\phi}}
\newcommand{\phidd}{\ddot{\phi}}

\newcommand{\eq}[1]{\begin{equation}{#1}\end{equation}}
\newcommand\nt{\addtocounter{equation}{1}\tag{\theequation}}
\newcommand{\mui}{^{\mu}}
\newcommand{\mli}{_{\mu}}
\newcommand{\nui}{^{\nu}}
\newcommand{\nli}{_{\nu}}
\newcommand{\pd}[2]{\frac{\partial {#1}}{\partial {#2}}}
\newcommand{\td}[2]{\frac{d{#1}}{d{#2}}}
\newcommand{\spd}[2]{\frac{\partial^2 {#1}}{\partial {#2}^2}}
\newcommand{\std}[2]{\frac{d^2 {#1}}{d{#2}^2}}

\newcommand{\tb}[1]{\textcolor{purple}{#1}}

\definecolor{dmgreen}{rgb}{0.90,0.50,0.10}
\newcommand{\dmc}[1]{\textcolor{dmgreen}{#1}}
\newcommand{\be}{\begin{equation}}
\newcommand{\ee}{\end{equation}}
%
%\section*{I. Introduction}

	\title{The rapid-turn inflationary attractor}
	\author{Theodor Bjorkmo}
	\email{t.bjorkmo@damtp.cam.ac.uk or theodor@bjorkmo.com}
	\affiliation{Department of Applied Mathematics and Theoretical Physics, DAMTP, University of Cambridge,
Cambridge, CB3 0WA, United Kingdom}

	\begin{abstract}
	\noindent
We prove the existence of a general class of rapidly turning two-field inflationary attractors. By only requiring a large, slowly varying turn rate, we solve the system completely without specifying any metric or potential, and prove the linear stability of the solution. Several recently studied turning inflation models, including hyperinflation, side-tracked inflation, and a flat field-space model, turn out to be examples of this general class of attractor solutions. Very rapidly turning models are of particular interest since they can be compatible with the swampland conjectures, and we show that the solutions further simplify in this limit.
	\end{abstract}
	
	\maketitle

\indent{\bf~Introduction --}
There has been much interest recently in multifield models of inflation with strongly non-geodesic motion \cite{Brown:2017osf,Mizuno:2017idt,Bjorkmo:2019aev,Christodoulidis:2018qdw,Cremonini:2010ua,Renaux-Petel:2015mga,Renaux-Petel:2017dia,Garcia-Saenz:2018ifx,Garcia-Saenz:2018vqf,Grocholski:2019mot, Fumagalli:2019noh,Roest:2019temp,Easson:2007dh,Achucarro:2010jv,Achucarro:2010da, Achucarro:2012sm,Achucarro:2012yr,Cespedes:2012hu,Hetz:2016ics,Chen:2018uul,Chen:2018brw}.
Such models can be realised in potentials that are far too steep for slow-roll, and have furthermore been shown to arise as natural attractor solutions in hyperbolic field-spaces \cite{Brown:2017osf,Mizuno:2017idt,Bjorkmo:2019aev,Christodoulidis:2018qdw,Renaux-Petel:2015mga,Renaux-Petel:2017dia,Garcia-Saenz:2018ifx,Garcia-Saenz:2018vqf,Grocholski:2019mot,Fumagalli:2019noh}, although they are by no means restricted to such geometries. In this paper we show that two-field models of inflation with a high turn rate can collectively be described by a single attractor solution, and that several notable examples of strongly turning inflation models are in fact examples of this attractor.

To find this attractor solution, we look for a general solution to the background equations of motion with  $\epsilon,|\eta|\ll1$ and a large, slowly varying turn rate $\omega$, meaning $\omega^2\gg\mathcal O(\epsilon)$ and $\nu\equiv \mathcal D_N\ln\omega\ll1$, %corresponding to an ``adiabaticity condition'' \cite{Cespedes:2012hu}.
since we want a stable, sustained phase of inflation.
We can do this without specifying any metric or potential by projecting the equations of motion onto a suitable vielbein basis, and the results are consequently completely general.
Examples of rapid-turn attractors include hyperinflation, side-tracked inflation, angular inflation, and the flat field-space turning inflation model of \cite{Achucarro:2012yr}, a diverse collection of models which demonstrates the universality of the solution.

%As we shall see, these conditions, together with $\epsilon$ and $\eta$ being small will be sufficient to allow us to completely solve the background equations of motion without specifying any potential or metric. Particular solutions can be then easily be recovered by evaluating the various projections of the covariant derivatives, and we give four examples that illustrate the power of this general approach.

With the background solution determined, one then immediately finds that two of the three independent components of the effective mass matrix of the perturbations are completely fixed (up to $\mathcal O(\epsilon)$ corrections). The equations of motion for the perturbations then only depend on two parameters, vastly simplifying the analysis. We will use this result to discuss the stability of the solution and the observables of these theories. We will also be interested in the behaviour of the solution in the limit $\omega\gtrsim180$, when it is compatible with the recently proposed swampland conjectures \cite{Garg:2018reu,Obied:2018sgi,Ooguri:2018wrx,Agrawal:2018own,ArkaniHamed:2006dz,Danielsson:2018ztv, Achucarro:2018vey,Kinney:2018nny}.
Finally we will show how particular solutions can easily be recovered by evaluating the various projections of the covariant derivatives of the potential, and we give four examples that illustrate the power of this general approach.

%In the limit where the turning rate gets very large, the solutions simplify even further. These very rapidly turning models are of particular interest since the conditions from the swampland conjectures \cite{Garg:2018reu,Obied:2018sgi,Ooguri:2018wrx,Agrawal:2018own,ArkaniHamed:2006dz,Danielsson:2018ztv} can be satisfied when the turn rate satisfies $\omega\geq180$ \cite{Achucarro:2018vey}.
%\tb{[Wrap up the intro better]}

%\section{II. Rapid-turn inflation}
\indent{\bf~Rapid-turn inflation --}
The analysis here rests on a vielbein formulation of inflation, using both the standard kinematic basis \cite{GrootNibbelink:2001qt,Peterson:2010np,Peterson:2011yt} and the potential gradient basis introduced in \cite{Bjorkmo:2019aev}. 

To solve the background evolution we want to find the (geometric scalar) field velocities $
\dot\phi_v\equiv v_a\dot\phi^a$ and $\dot\phi_w\equiv w_a\dot\phi^a$ where $v_a=V_{;a}/\|V_{;b}\|$ and $w_a$ is a (co)vector field orthonormal to $v^a$. The equations of motion for these velocities are given by \cite{Bjorkmo:2019aev}
\begin{align}
\ddot\phi_v&=-3H\dot\phi_v-\Vv +\Omega_v\dot\phi_w\label{eq:veq}\\
\ddot\phi_w&=-3H\dot\phi_w-\Omega_v\dot\phi_v \, ,\label{eq:weq}
\end{align}
where $\Vv =v^aV_{;a}$, $\Vvw =v^aw^bV_{;ab}$ etc, $\Omega_v\equiv w_a\mathcal D_t v^a=(\Vvw \dot\phi_v+\Vww \dot\phi_w)/\Vv $, and $\mathcal D_t X^a\equiv\dot X^a+\Gamma^a_{bc}X^b\dot\phi^c$. We also define $\mathcal D_{N}\equiv H^{-1}\mathcal D_t$.

The reason for using the gradient basis $e_I^a=(v^a,w^a)$ above is that unlike in the kinematic basis $e_I^a=(n^a,s^a)$, where $n^a=\dot\phi^a/\dot\phi$ and $s^a=(-\dot\phi_wv^a+\dot\phi_vw^a)/\dot\phi$,
we already know the directions of $v^a$ and $w^a$ before we have the full solution, making them much easier to work with. In this paper we are looking at inflation models with a significant turn rate, meaning $\omega^2\gg\mathcal O(\epsilon)$. The turn rate $\omega\equiv s_a\mathcal D_N n^a$ will be a crucial quantity to our analysis, and using the gradient basis it may be expressed as
\eq{
\omega={\dot\phi_w\Vv }/{H\dot\phi^2},\label{eq:omegadef}
}
which is obtained using the Klein-Gordon equation \cite{Bjorkmo:2019aev}.

However, while the gradient basis is very useful for solving the background equations of motion, when dealing with perturbations the kinematic basis has significant advantages. This is because we can automatically see which perturbations are adiabatic and which correspond to isocurvature modes. %Defining $\mathcal D_N\ln\omega\equiv\nu$ and $\delta\pi^I\equiv d\delta\phi^I/dN$, the equations of motion for the perturbations are given by
%\eq{
%\begin{pmatrix}\delta\phi_n' \\ \delta\phi_s' \\ \delta\pi_n' \\ \delta\pi_s'\end{pmatrix}=
%\begin{pmatrix}0 & 0 & 1 &0\\
%0 & 0 & 0 &1\\
%\frac{k^2}{a^2H^2}-\Mnn+\omega^2 &-\Mns+\omega(3-\epsilon+\nu)&-3+\epsilon & 2\omega\\
%-\Mns-\omega (3-\epsilon+\nu)&\frac{k^2}{a^2H^2} -\Mss+\omega^2& -2\omega & -3+\epsilon
%\end{pmatrix}
% \begin{pmatrix}\delta\phi_n \\ \delta\phi_s \\ \delta\pi_n \\ \delta\pi_s\end{pmatrix},\label{eq:evo}
%}
For compactness denoting $d/dN$ by $'$s, the equations of motion for the perturbations can be written \cite{Achucarro:2010da,GrootNibbelink:2001qt,Sasaki:1995aw}
\eq{
\delta\phi''^I+[(3-\epsilon)\delta^I_J-2\omega\epsilon^I_J]\delta\phi'^J+C(k)^I_J\delta\phi^J=0\label{eq:evo}
}
where $\epsilon^1_2=-\epsilon^2_1=1$ and $C(k)^I_J$ is given by
\eq{
\hspace{-0.12cm}C(k)^I_J=\begin{pmatrix}
\Mnn -\omega^2+\frac{k^2}{a^2H^2} &\Mns-\omega(3-\epsilon+\nu)\\
\Mns+\omega (3-\epsilon+\nu)&\Mss-\omega^2+\frac{k^2}{a^2H^2}
\end{pmatrix},
}
where we further defined $\Mnn=n^an^bM_{ab}/H^2$, $\Mns=n^as^bM_{ab}/H^2$, and $\Mss=s^as^bM_{ab}/H^2$ as projections of the dimensionless mass matrix. These are given by
\begin{align*}
\Mnn &=\frac{\Vvv \dot\phi_v^2+2\Vvw \dot\phi_v\dot\phi_w+\Vww \dot\phi_w^2}{H^2\dot\phi^2}-2\epsilon(3-\epsilon+\eta)\\
\Mns&=\frac{(\Vww -\Vvv )\dot\phi_v\dot\phi_w+\Vvw (\dot\phi_v^2-\dot\phi_w^2)}{H^2\dot\phi^2}-2\omega\epsilon\label{eq:massmatrix}\nt\\
%\Mss
\Mss&=\frac{\Vww \dot\phi_v^2-2\Vvw \dot\phi_v\dot\phi_w+\Vvv \dot\phi_w^2+R\dot\phi^4/2}{H^2\dot\phi^2},
\end{align*}
where in the middle line we used the definition of the turning rate to tidy up the expression.

%\subsection{Finding the attractor solution}
\indent{\it~Finding the attractor solution --}
During inflation we need our field velocities $\dot\phi_v$ and $\dot\phi_w$ to satisfy $\ddot\phi_i\sim\mathcal O(\epsilon)H\dot\phi_i$, which is necessary for $\epsilon$ and $\eta$ to be small. The same goes for the total field velocity, $\mathcal D_t \dot\phi^2=\mathcal O(\epsilon)H\dot\phi^2$, and so the equations of motion for $\dot\phi_v$ and $\dot\phi_w$ together give
\eq{
\dot\phi_v\Vv/H\dot\phi^2=-3+\mathcal O(\epsilon).
}
With our expression for $\omega$ in equation \ref{eq:omegadef} this immediately implies
\eq{
\frac{\dot\phi_v}{\dot\phi}=\frac{-3}{\sqrt{9+\omega^2}},\quad\frac{\dot\phi_w}{\dot\phi}=\frac{\omega}{\sqrt{9+\omega^2}} .\label{eq:velocities}
}
which is accurate when $\omega^2\gg\mathcal O(\epsilon)$. It also follows that
\eq{
\dot\phi=\frac{\Vv }{H\sqrt{\omega^2+9}}\label{eq:dotphisq}.
}
%When the turn rate is large, one can thus have small value for $\epsilon$ even if $\epsilon_v$ is large.
As we can see, in this vielbein basis, the field velocities are automatically fixed by the turn rate $\omega$. The turn rate is still an unknown, but ensuring that the field equations of motion are statisfied individually and that $\omega$ varies slowly will allow us to fix it, and thus find a complete solution. As a brief aside, we also note that the above equation implies $\epsilon_v=\epsilon(1+\omega^2/9)$  \cite{Hetz:2016ics,Achucarro:2018vey}, which is why rapid turn inflation can be realised in potentials that are too steep for slow-roll.

Looking at the equation of motion for $\dot\phi_v$ in \ref{eq:veq} alone, and using the results in equations \ref{eq:velocities} and \ref{eq:dotphisq} we can now deduce $\Omega_v/H=\omega+\mathcal O(\epsilon/\omega)$, from which one can show
\eq{
\frac{\Vww}{H^2}-\frac{3}{\omega}\frac{\Vvw}{H^2}=\omega^2+9+\mathcal O(\epsilon)\label{eq:omega1}.
}
The equation of motion for $\dot\phi_w$ requires $\Omega_v/H=\omega+\mathcal O(\epsilon)$, which is compatible with the former (but less stringent). Intuitively, requiring $\Omega_v/H\simeq\omega$ makes sense: equation \ref{eq:velocities} fixes what fractions of the field velocity are parallel and orthogonal to the gradient direction, and if this is to be maintained as the fields turn, the basis vectors need to turn at the same rate.

Finally, we want this solution to have a slowly varying turn rate. The next step is therefore to compute $\nu=\mathcal D_N\ln\omega$, and to do so we will use the expression for $\omega$ in equation \ref{eq:omegadef}. From this we see that $\nu=\mathcal O(\epsilon)$ requires $\mathcal D_N\ln \Vv=\mathcal O(\epsilon)$, which implies
\eq{
\frac{\Vvw}{H^2}-\frac{3}{\omega}\frac{\Vvv}{H^2}=\mathcal O(\omega\epsilon),\label{eq:omega2}
}
where we used equations \ref{eq:velocities} and \ref{eq:dotphisq}. These two equations can also be combined into the convenient form
\eq{
\frac{\Vww}{H^2}-\frac{9}{\omega^2}\frac{\Vvv}{H^2}=\omega^2+9+\mathcal O(\epsilon)\label{eq:omega3}.
}

When $\Vvv/H^2\lesssim\mathcal O(\omega^2\epsilon)$ and $\Vvw/H^2\lesssim\mathcal O(\omega\epsilon)$, one straightforwardly finds that the turn rate is given by $\omega^2\simeq{\Vww}/{H^2}-9$. However, when $\Vvv$ and $\Vvw$ are non-negligible, the situation is a little bit more complicated. The direction of motion must now be chosen so that $\dot \Vv=\Vvv \dot\phi_v+\Vvw\dot\phi_w\approx 0$, imposing a second constraint on the turn rate beyond ensuring that the equations of motion are satisfied. In this scenario one can find two particularly convenient expressions for $\omega$,
\eq{
\omega=\frac{3\Vvv}{\Vvw},\hspace{0.5cm}
\omega^2=\frac{\Vww }{H^2}-\frac{\Vvw^2}{\Vvv^2}\frac\Vvv{H^2}-9,
}
which must be matched if rapid-turn inflation is to take place. This is generally not possible everywhere in the target space, and will restrict where rapid-turn inflation may happen. Examples of this will be given later.

A technical detail that should be mentioned is that what we have really done is to find the leading order parts of the solutions $\dot\phi_v=\db\phi_v+\delta\dot\phi_v$ and $\dot\phi_w=\db\phi_w+\delta\dot\phi_w$ that we take to be functions of field-space position only. This gives us an approximate solution, which is valid as long as the necessary corrections are small, i.e. $\delta\dot\phi_i\sim \mathcal O(H\db\phi_i \epsilon)$. For this to be the case, the explicit time derivatives of the field velocities must satisfy $d\db\phi_i/dt\sim O(H\db\phi_i \epsilon)$, which requires the terms $\Vv$, $\Vvw$, etc to vary slowly along the trajectory. This is a very much consistent with our assumption of $\nu=\mathcal O(\epsilon)$.

%\subsection*{Stability}
\indent{\it~Perturbations and stability --}
Now we would like to look at the behaviour of perturbations in these models. We first need to verify that the background solution is stable, a necessary condition for it to be an attractor. To begin, we note that the Klein-Gordon equation gives
\eq{
\mathcal D_N\omega=-\Mns+\omega(-3+\epsilon-\eta)\label{eq:omeganu}
}
Demanding $\nu=\mathcal D_N\ln\omega\sim\mathcal O(\epsilon)$ therefore requires
\eq{
\Mns=-3\omega+\mathcal O(\omega\epsilon)\label{eq:Mnscond},
}
and we have fixed one element of the effective mass matrix. One can also show that equations \ref{eq:omega2} and \ref{eq:omega3} imply
\eq{
\Mnn=\omega^2+\mathcal O(\epsilon)\label{eq:Mnncond}.
}
We now define $\delta\pi_I\equiv\delta\phi_I'$, and find that the equations of motion for the perturbations can be written
\eq{
\begin{pmatrix}\delta\phi_n' \\ \delta\phi_s' \\ \delta\pi_n' \\ \delta\pi_s'\end{pmatrix}\hspace{-0.1cm}=\hspace{-0.1cm}
\begin{pmatrix}0 & 0 & 1 &0\\
0 & 0 & 0 &1\\
-\kappa^2 &6\omega&-3& 2\omega\\
0 &\omega^2-\Mss-\kappa^2& -2\omega & -3
\end{pmatrix}\hspace{-0.1cm}
 \begin{pmatrix}\delta\phi_n \\ \delta\phi_s \\ \delta\pi_n \\ \delta\pi_s\end{pmatrix}\label{eq:perteq}
}
where $\kappa\equiv k/aH$ and we have ignored $\mathcal O(\epsilon)$ corrections.

To see whether the background solution is stable, we look at the eigenvalues of the evolution matrix (the local Lyapunov exponents) when $k=0$. They are given by
\eq{
\lambda=-3,\quad0,\quad\frac12\left(-3\pm\sqrt{9-4\Mss-12\omega^2}\right).
}
As long as the dimensionless entropic mass satisfies $\Mss>-3\omega^2$, the system has one (near) constant mode (the adiabatic one) and three decaying ones, and is thus stable. Using equations \ref{eq:massmatrix} and \ref{eq:omega1} one can show that the entropic mass can be written in the convenient form
\eq{
\Mss=\frac{9 \Vww }{H^2(\omega^2+9)}+\frac{\Vvv }{H^2}+\frac{R\dot\phi^2}{2H^2}+\mathcal O(\epsilon)\label{eq:tMss}
}
which must satisfy the above condition.

Notably, equation \ref{eq:perteq} only depends on two free paramters, $\omega$ and $\mu_s$, which have a significant impact on the primordial perturbations generated by these models. If $\xi\equiv\Mss/\omega^2>1$, the perturbations can be analysed using a single-field EFT with a reduced speed of sound, which has been studied extensively in the literature (see e.g. \cite{Achucarro:2010jv,Achucarro:2010da,Achucarro:2012sm,Achucarro:2012yr,Cespedes:2012hu}). Observationally, these models are generally characterised by large equilateral non-Gaussianity ($\propto c_s^{-2}$), and a suppressed tensor-scalar ratio ($\propto c_s)$. If $\xi<1$, the situation is rather different. One can show, by looking at the evolution equations for $v^I=a\delta\phi^I$, that the perturbations have a growing mode in the last $\ln(\sqrt{1-\xi}\omega)$ e-folds before horizon crossing. Integrating this eigenvalue, one can approximate the sub-horizon growth of the power spectrum, $\gamma(\omega,\xi)^2$, which is found to be given by \cite{future}
\eq{
\ln(\gamma^2)\approx(2-\sqrt{3+\xi})\pi\omega.
}
It was recently shown \cite{Fumagalli:2019noh} that these models are well described by a single-field EFT with an imaginary speed of sound, which generically predicts enhanced non-Gaussianity in flattened configurations. However this EFT cannot itself predict the growth of the power spectrum, given here, which naturally leads to a suppression of the tensor-scalar ratio. Interestingly, it was found that if the growth is too large, perturbative control is lost. To limit the growth, we either require $\omega\lesssim\mathcal O(10)$ or $\xi$ very close to 1, restricting the observationally allowed parameter space. A detailed discussion of this is however beyond the scope of this paper.

%\section{III. The swampland limit: $\omega\geq180$}
\indent{\it~Rapid-turn inflation and the swampland%The swampland limit: $\omega\geq180$
~--}
For inflation to be compatible with the swampland conjectures
\cite{Garg:2018reu,Obied:2018sgi,Ooguri:2018wrx,Agrawal:2018own,ArkaniHamed:2006dz,Danielsson:2018ztv},
 the field excursion must be bounded, $\Delta\phi\leq\Mp\equiv(8\pi G)^{-1/2}$, which implies $\epsilon\leq1/2N^2$ (where $N$ is the number of e-folds of inflation), and either the gradient satisfies $\Mp\Vv/V\geq c$ or the minimum eigenvalue of the Hessian satisfies $\Mp^2\hspace{1pt}\text{min}(V_{;ab})/V\leq-c'$, where $c$ and $c'$ are $\mathcal O(1)$ coefficients.

If we want to satisfy the swampland conjectures by having a large gradient, the turn rate will satisfy $\omega\gtrsim180$ \cite{Achucarro:2018vey}. In this limit, the rapid-turn attractor solution drastically simplifies, assuming that $\Vvv$ is not parametrically larger than $\Vww $, and we find
\eq{
\omega^2\simeq\frac{{\Vww }}{H^2},\quad\dot\phi_v\simeq-\frac{3H\Vv }{\Vww },\quad \dot\phi^2\simeq\frac{\Vv ^2}{\Vww }.\label{eq:veryrapid}
}
The two conditions can then be reformulated as
\eq{
{\Mp \Vv }/{V}\geq c,\quad{\Mp \Vww }/{\Vv }\geq 3c N^2,
}
which are straightforward to check.

If instead we want to satisfy the condition on the Hessian, there are fewer simplifications that can be made, since the turn rate is no longer necessarily very large.

For a given model, we do not have full freedom in choosing which of these conditions we want to satisfy, and there are pitfalls that need to be avoided. The spectral index of the power spectrum needs to be matched with obserervations, which constrains the background solution, potentially ruling out one or both of these options (c.f. \cite{Bjorkmo:2019aev}). Moreover, as discussed earlier, in very rapidly turning models ($\omega\geq\mathcal O(100)$), entropic masses $\mu_s<\omega^2$ can cause loss of perturbative control over the primordial perturbations. This is especially problematic for models realised in hyperbolic field-spaces, where $R<0$. From equations \ref{eq:tMss} and \ref{eq:veryrapid} one can deduce that these models need at least $\Vvv\gtrsim\Vww$ to avoid this problem, which in turn implies that these models will be of the type where both the constraints in equations \ref{eq:omega2} and \ref{eq:omega3} are important.

%\section*{IV. Examples of rapid-turn attractors}
\indent{\bf~Examples of rapid-turn attractors --}
In this section we show that several non-standard two-field inflation models are examples of rapid-turn attractors, and use the relations derived above to straightforwardly find the form of the solutions. Hyperinflation and the turning inflation model of \cite{Achucarro:2012yr} present two algebraically simple models where we can easily derive the form of the full solution, and for the latter we also give some numerical examples of the agreement between the attractor model predictions and numerical simulation. Side-tracked inflation and angular inflation, however, are algebraically messy. In the former case we present more numerical examples of the agreement between the attractor predictions and simulations, and in the latter case we show how we can straightforwardly recover the parametric equation for the field trajectory.

The point of this section is not to delve into the details of these models; it is instead to show what a broad class of models can be described by this attractor solution, and to demonstrate how powerful these techniques can be for finding explicit solutions given a metric and potential. In models where $\Vvv$ and $\Vvw $ are negligible (e.g. hyperinflation), rapid-turn inflation can occur anywhere in the target space, and equations \ref{eq:dotphisq} and \ref{eq:omega3} immediately give the full solution (i.e. $\dot\phi_v$ and $\dot\phi_w$) which is valid everywhere. In models where $\Vvv$ and $\Vvw $ are non-negligible, however, the expressions for $\omega$ in equations \ref{eq:omega2} and \ref{eq:omega3} must be matched, and solving these two equations will tell us where in field-space this phase may occur.

%\subsection*{Hyperinflation}
\indent{\it~Hyperinflation --}
This model, which was first introduced by \cite{Brown:2017osf}, and then further studied in \cite{Mizuno:2017idt,Bjorkmo:2019aev}, provides a very clean example of a rapid-turn attractor. The usual metric and potential used for this model are
\eq{
ds^2=d\phi^2+L^2\sinh(\phi/L)^2d\theta^2,\quad V=V(\phi),
}
although more generally it just requires $\Vww =\Vv /L$, $\Vvw \simeq0$ and $\Vvv \ll \Vww $. Hyperinflation refers to a new phase of inflation that occurs when the potential becomes sufficiently steep for slow-roll to be geometrically destabilised \cite{Brown:2017osf,Renaux-Petel:2015mga}. Using the results derived earlier in this paper and the above covariant derivatives of the potential, one immediately finds
\eq{
\dot\phi_v=-3HL,\quad \dot\phi^2=L\Vv ,
}
which is only a consistent solution if $L\Vv >9H^2L^2$, which is precisely the condition for slow-roll to become unstable \cite{Brown:2017osf,Renaux-Petel:2015mga}. Its entropic mass, given by $\Mss=-\omega^2+\mathcal O(\epsilon)$, is very tachyonic in the $\omega\gg1$ limit, but the background evolution is nevertheless stable.

\begin{figure}
    \centering
           \begin{subfigure}{0.41\textwidth}
         \includegraphics[width=1\textwidth]{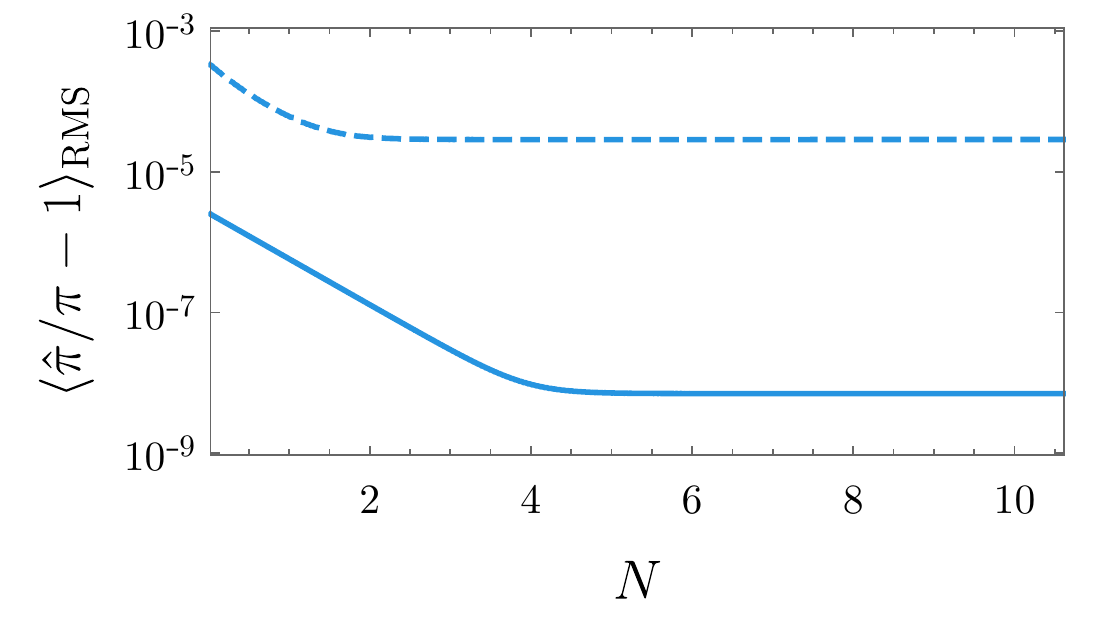}
   % \caption{Fraction of negative eigenvalues}
%    \label{fig:diso}
    \end{subfigure}
%    \caption{These graphs show the effect of varying the }\label{fig:dSHiso}
%\end{figure}
%
%\begin{figure}
%    \centering
%~
%	    \begin{subfigure}{0.41\textwidth}
%         \includegraphics[width=1\textwidth]{turninginflationpiwRMS.pdf}
%   % \caption{e-folds before end when a second direction becoms tachyonic}
%%    \label{fig:dns}
%    \end{subfigure}
    \vspace{-0.15cm}
     %add desired spacing between images, e. g. ~, \quad, \qquad, \hfill etc. 
      %(or a blank line to force the subfigure onto a new line) \\
	    \caption{RMS fractional differences over a few oscillation cycles between the predicted $\hat\pi_v$ (dashed), $\hat\pi_w$ (solid) and numerically calculated field momenta $\pi_v$, $\pi_w$ in the flat field-space model as the solutions converge. \label{fig:turning}}
\end{figure}

%\subsection*{Flat field-space model}
\indent{\it~A flat field-space model --}
This example, first introduced in \cite{Achucarro:2012yr}, is very different from hyperinflation, especially since the field space is not hyperbolic. Here we have the metric and potential
\eq{
ds^2=d\rho^2+\rho^2d\theta^2,~ V=V_0-\alpha\theta+\frac12m^2(\rho-\rho_0)^2.
}
In this model, the rapid-turn regime appears at $\rho\gg\rho_0$, where the gradient of the potential is dominated by the $\rho$-direction, but where the $\theta$-term nevertheless plays an important role. Here we have $\Vv \simeq m^2\rho$, $\Vvv \simeq \Vww \simeq m^2$, and $\Vvw \simeq{\alpha}/{\rho^2}$. Equations \ref{eq:omega2} and \ref{eq:omega3} thus give us two expressions for $\omega$:
\eq{
\omega%=\frac{3\Vvv }{\Vvw }
\simeq{3\rho^2m^2}/{\alpha},\quad \omega%=\frac{\sqrt{\Vww }}H
\simeq{\sqrt{3}\Mp m}/{\sqrt{V_0}}.
}
These must match (up to a sign), telling us that this type of inflation can only happen at
\eq{
\rho^2={\Mp\alpha}/{\sqrt{3V_0}m},\quad\Rightarrow\quad\dot\theta={\dot\phi_w}/{\rho}\simeq m,
}
in agreement with \cite{Achucarro:2012yr}. Moreover, to illustrate the accuracy of the rapid-turn attractor predictions, Figure \ref{fig:turning} shows the agreement between predictions and simulations for a swampland compatible model with $\alpha=5.0 \times10^{-16}$, $m=2.5\times 10^{-3}$ and $V_0=3.4\times 10^{-10}$, giving $\omega\approx 230$.

%\begin{figure}
%    \centering
%           \hspace{-0.1cm}\begin{subfigure}{0.48\textwidth}
%         \includegraphics[width=1\textwidth]{turninginflationpiw.pdf}
%   % \caption{Fraction of negative eigenvalues}
%%    \label{fig:diso}
%    \end{subfigure}
%%    \caption{These graphs show the effect of varying the }\label{fig:dSHiso}
%%\end{figure}
%%
%%\begin{figure}
%%    \centering
%\\
%%
%	    \hspace{1cm}\begin{subfigure}{0.42\textwidth}
%         \includegraphics[width=1\textwidth]{turninginflationpiwdiff.pdf}
%   % \caption{e-folds before end when a second direction becoms tachyonic}
%%    \label{fig:dns}
%    \end{subfigure}
%     %add desired spacing between images, e. g. ~, \quad, \qquad, \hfill etc. 
%      %(or a blank line to force the subfigure onto a new line) \\
%	    \caption{The momentum component $\pi_w\equiv\dot\phi_w$ in the flat field space example example. In the first graph, the attractor model prediction (green) completely covers the numerical value (blue). The second graph plots $\|\hat\pi_w/\pi_w-1\|$, where $\hat\pi_w$ is the predicted value from equation \ref{eq:omegasqexpr}. \label{fig:turning}}
%\end{figure}

\begin{figure}
    \centering
        \begin{subfigure}{0.39\textwidth}
        \includegraphics[width=1\textwidth]{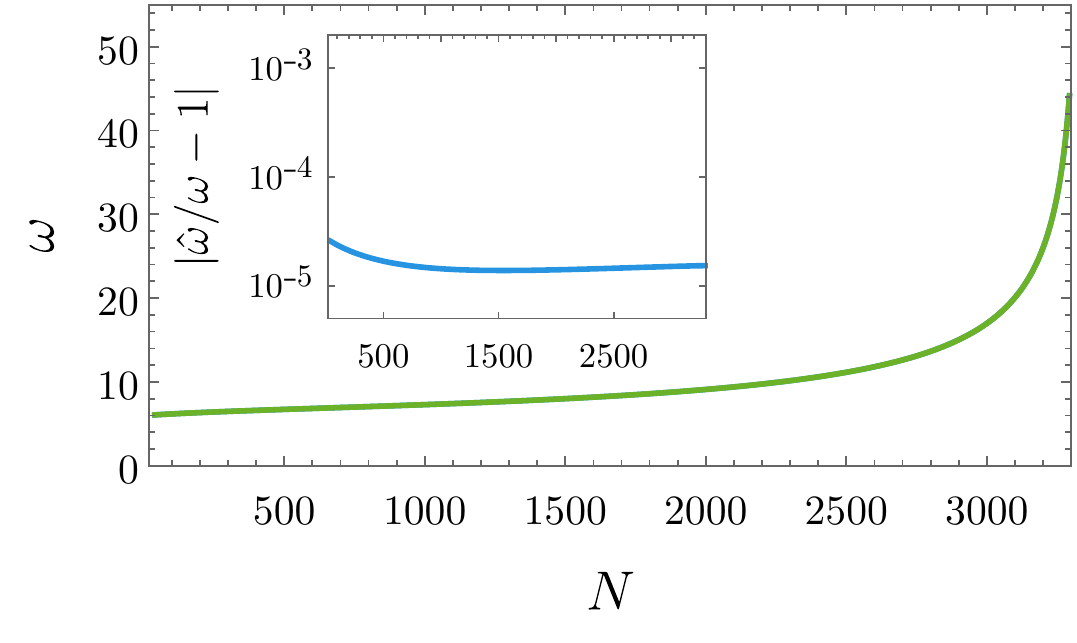}
   % \caption{Fraction of negative eigenvalues}
%    \label{fig:diso}
    \end{subfigure}
%    \caption{These graphs show the effect of varying the }\label{fig:dSHiso}
%\end{figure}
%
%\begin{figure}
%    \centering
%~
%%
%    \begin{subfigure}{0.41\textwidth}
%         \includegraphics[width=1\textwidth]{sidetrackedomegadiff.pdf}
%   % \caption{e-folds before end when a second direction becoms tachyonic}
%%    \label{fig:dns}
%    \end{subfigure}
%     %add desired spacing between images, e. g. ~, \quad, \qquad, \hfill etc. 
%      %(or a blank line to force the subfigure onto a new line) \\
          \vspace{-0.15cm}
	    \caption{The turn rate $\omega$ in the side-tracked example (green), and the fractional difference between the numerical and predicted values (blue). \label{fig:sidetracked}}
\end{figure}

%\subsection*{Side-tracked inflation}
\indent{\it~Side-tracked inflation --}
Side-tracked inflation is another example of a rapid-turn attractor in hyperbolic geometry, which like hyperinflation can arise after geometric destabilisation of slow-roll \cite{Renaux-Petel:2015mga,Garcia-Saenz:2018ifx}. Looking at the side-tracked inflation model with the so called `minimal geometry', we have
\eq{
ds^2=\left(1+\frac{2\chi^2}{M^2}\right)d\phi^2+d\chi^2,~ V=U(\phi)+\frac{m_h^2}2\chi^2.
}
In this model, $m_h$ is the mass of a heavy field with $m_h\gg H$, but despite the size of this mass, slow-roll is destabilised by the negative curvature, and we end up in a `side-tracked' phase of inflation, which is another example of a rapid-turn attractor. $\Vvv $ and $\Vvw $ are non-negligible in this model, so we can use equations \ref{eq:omega2} and \ref{eq:omega3} to find where in field-space side-tracked inflation may happen. Assuming $MU_{,\phi\phi}\ll U_{,\phi}$ \cite{Garcia-Saenz:2018ifx} and that $U$ dominates the potential energy, we find that this happens at
\eq{
\frac{2\chi^2}{M^2}=\sqrt{\frac23}\frac{\Mp|U_{,\phi}|}{m_hM\sqrt{U}}-1,
}
recovering the expression found in \cite{Garcia-Saenz:2018ifx}. Figure \ref{fig:sidetracked} also illustrates the agreement between the rapid-turn attractor predictions and numerical simulations for a model with a natural inflation potential $U(\phi)$ with $M=0.001\Mp$ and $m_h/H=10$.

%\subsection*{Angular inflation}
\indent{\it~Angular inflation --}
An additional attractor model was found recently by \cite{Christodoulidis:2018qdw}, in the context of $\alpha$-attractor models \cite{Kallosh:2013hoa,Kallosh:2013yoa,Kallosh:2013yoa,Kallosh:2014rga,Linde:2015uga,Carrasco:2015rva,Carrasco:2015uma,Kallosh:2017wku,Achucarro:2017ing}, where the geometry again is hyperbolic. Here we work with a metric and potential of the form
\eq{
ds^2=\frac{6\alpha(d\phi^2+d\chi^2)}{(1-\phi^2-\chi^2)^2},\quad V=\frac{\alpha}6\left(m_\phi^2\phi^2+m_\chi^2\chi^2\right).
}
Reparametrising this as $\phi=r\cos\theta$, $\chi=r\sin\theta$, and defining $R=m_\chi^2/m_\phi^2$, they found a new angular attractor solution at $1-r\ll1$ when the parameters $\alpha$ and $R$ satisfied $\alpha\ll1$ and $R\gtrsim10$. Defining $\delta\equiv1-r^2$, one can then solve equations \ref{eq:omega2} and \ref{eq:omega3} (eliminating $\omega$) to leading order in $\delta$, to find its parametrisation by $\theta$ as written in \cite{Christodoulidis:2018qdw}\footnote{The current arXiv version (v1) has a typo in equation (2.14), where $\cot$ and $\tan$ appear with squares. If (2.13) is expanded in $\alpha$, however, these squares do not appear.}:
\eq{
\delta(\theta)=1-r(\theta)^2=\frac{9\alpha(\cot\theta+R\tan\theta)^2}{2(R-1)^2}.
}

\indent{\bf~Conclusions --} We have shown that there exists a completely general rapidly turning attractor solution in two-field inflation. The attractor is not restricted to any particular background geometry or form of the potential, and we have shown how several recently studied non-standard inflationary attractors are in fact examples of this rapid-turn attractor.

Having a large, slowly varying turn rate was sufficient to allow the equations of motion to be solved in generality. Moreover, we showed that these solutions have two out of three elements in the effective mass matrix constrained up to $\mathcal O(\epsilon)$ corrections. Only the turn rate and entropic mass remain unconstrained degrees of freedom. The primordial perturbations generated by these theories can therefore be expected to be determined by these parameters, simplifying future analyses of these models.

The analysis here has been restricted to two fields, but we expect it to generalise to models with more than two fields. Such solutions have been explicitly shown to exist in the context of hyperinflation \cite{Bjorkmo:2019aev}, but a general analysis is rather non-trivial and is left for future work.

\indent{\bf~Acknowledgements --}
I am grateful for stimulating and helpful discussions with Anne-Christine Davis, Bogdan Ganchev, David Marsh, and Sebastien Renaux-Petel. I am supported by an STFC studentship.

\bibliography{RTIrefs}

%\begin{thebibliography}{99}

%\bibitem{a}
%Author, \emph{Title}, \emph{J. Abbrev.} {\bf vol} (year) pg.
%
%\bibitem{b}
%Author, \emph{Title},
%arxiv:1234.5678.
%
%\bibitem{c}
%Author, \emph{Title},
%Publisher (year).

% Please avoid comments such as "For a review'', "For some examples",
% "and references therein" or move them in the text. In general,
% please leave only references in the bibliography and move all
% accessory text in footnotes.

% Also, please have only one work for each \bibitem.

%\end{thebibliography}
\end{document}